\documentclass[aps,showpacs,prl,10pt,twocolumn,superscriptaddress,A4]{revtex4-1}
\usepackage[pdftex]{graphicx}
\DeclareGraphicsExtensions{.pdf}

\usepackage{amsmath}
\usepackage{amssymb}
\usepackage{amscd}
\usepackage{wasysym}
\usepackage[ansinew]{inputenc}
\usepackage[T1]{fontenc}
\usepackage{ae,aecompl}
\usepackage{hyperref}

\usepackage{color}
\definecolor{red}{rgb}{1,0,0}
\definecolor{blue}{rgb}{0,0,0.8}
\definecolor{green}{rgb}{0,0.5,0}
\definecolor{lred}{rgb}{1,0.7,0.7}
\definecolor{lgreen}{rgb}{0.7,1,0.7}
\definecolor{llred}{rgb}{0.98,0.98,0.98}
\definecolor{llgreen}{rgb}{1,1,1}
\usepackage{colortbl}

\newcommand{\be}{\begin{equation}}
\newcommand{\ee}{\end{equation}}
\newcommand{\bea}{\begin{eqnarray}}
\newcommand{\eea}{\end{eqnarray}}

\newcommand{\revision}[1]{\textcolor{black}{#1}}

\makeatletter
\begin{document}
\bibliographystyle{apsrev}

\title{Escape Routes, Weak Links, and Desynchronization in Fluctuation-driven Networks}

\author{Benjamin Sch\"afer}
\thanks{These authors contributed equally.}
\affiliation{Network Dynamics, Max-Planck-Institute for Dynamics and Self-Organization (MPI DS), 37077 G\"ottingen, Germany}

\author{Moritz Matthiae}
\thanks{These authors contributed equally.}
\affiliation{Network Dynamics, Max-Planck-Institute for Dynamics and Self-Organization (MPI DS), 37077 G\"ottingen, Germany}
\affiliation{Forschungszentrum J\"ulich, Institute for Energy and Climate Research -
	Systems Analysis and Technology Evaluation (IEK-STE),  52428 J\"ulich, Germany}
\affiliation{Department of Micro- and Nanotechnology, Technical University of Denmark, 2800 Kongens Lyngby, Denmark}

\author{Xiaozhu Zhang}
\affiliation{Network Dynamics, Max-Planck-Institute for Dynamics and Self-Organization (MPI DS), 37077 G\"ottingen, Germany}

\author{Martin Rohden}
\affiliation{Network Dynamics, Max-Planck-Institute for Dynamics and Self-Organization (MPI DS), 37077 G\"ottingen, Germany}
\affiliation{Jacobs University, Department of Physics and Earth Sciences, 28759 Bremen, Germany}

\author{Marc Timme}
\affiliation{Network Dynamics, Max-Planck-Institute for Dynamics and Self-Organization (MPI DS), 37077 G\"ottingen, Germany}
\affiliation{Department of Physics, Technical University of Darmstadt, 64289 Darmstadt, Germany}
\affiliation{Institute for Theoretical Physics, Technical University of Dresden, 01062 Dresden Germany}

\author{Dirk Witthaut}
\affiliation{Forschungszentrum J\"ulich, Institute for Energy and Climate Research -
	Systems Analysis and Technology Evaluation (IEK-STE),  52428 J\"ulich, Germany}
\affiliation{Institute for Theoretical Physics, University of Cologne, 
		50937 K\"oln, Germany}

\date{\today }

\begin{abstract}
Shifting our electricity generation from fossil fuel to renewable energy sources introduces large fluctuations to the power system. Here, we demonstrate how increased fluctuations, reduced damping and reduced intertia may undermine the dynamical robustness of power grid networks. Focusing on fundamental noise models, we derive analytic insights into which factors limit the dynamic robustness and how fluctuations may induce a system escape from an operating state. Moreover, we identify weak links in the grid that make it particularly vulnerable to fluctuations. These results thereby not only contribute to a theoretical understanding of how fluctuations act on distributed network dynamics, they may also help designing future renewable energy systems to be more robust.
\end{abstract}


\maketitle

% --- content ------------------------------------------------

The development of a sustainable energy supply is one of the major technical, economical and societal challenges of the coming decades.  To mitigate climate change, the majority of existing fossil-fuel power plants will be replaced by renewable energy sources, in particular wind and solar power \cite{Amin05,Bruc14}. This requires a comprehensive reengineering of the electric power grid as well as a better understanding of the operation, dynamics and stability of complex networked systems \cite{Stro01,Marr08,Brum13,15editorial}.

The operation of wind turbines and photovoltaics is essentially determined by the weather such that the power generation fluctuates strongly on all time scales \cite{Sims11}. Large storage and backup infrastructures are needed to balance power generation and demand on time scales of hours to months \cite{Heid10}. In addition, the high-frequency fluctuations can be enormous, in particular due to the turbulent character of wind power \cite{Mila13,Anva16}. At the same time, large fossil-fuel plants are closed down such that the effective inertia and damping decreases, making the power grid more vulnerable to instabilities by transients \cite{Ulbig2014}.

In this Letter we analyze how high-frequency fluctuations threaten the dynamical robustness of electric power grids. The stable operation of a grid requires all generators and loads to remain phase-locked. We analyze the robustness of this phase-locked state, mapping the grid dynamics to an escape problem in a high-dimensional stochastic dynamical system. This analysis reveals essential factors which limit the operability of future power grids. Furthermore, we uncover how the network topology determines the robustness of the grid and identify potential escape routes and vulnerable links in complex networks. 

\textit{Synchronization and robustness of electric power grids.--}
Consider first the dynamics of coupled synchronous generators and consumers $j \in \{1,\ldots,N\}$. Each unit's dynamics is described by the swing equation \cite{Mach08,Fila08,12powergrid} for the mechanical rotor angle $\delta_j$ and the phase velocity $\omega_j$ relative to the grid reference frequency $\Omega = 2\pi \times 50 \, (\mbox{or} \, 60)$ Hz,
\begin{align}
\dot \delta_j &= \omega_j \\
\frac{2H_j}{\Omega} \dot \omega_j + 2 \Omega  D_j \omega_j &= P_j - P^{\rm (el)}_j,
\label{eqn:swing}
\end{align}
where the right hand side is the net power acting on the machine. The swing equation is made dimensionless by dividing all terms by the rated power of the machine, which is referred to as the `per unit system' in engineering \cite{Mach08}. The inertia constant $H_j$ then gives the stored kinetic energy at synchronous speed which is proportional to the moment of inertia of the $j$th machine and $D_j$ is a damping constant.  If not mentioned otherwise, we assume a typical value of inertia $H_j=H=4 \, {\rm s}$ and a damping constant of $D_j=D = 4\cdot 10^{-5}\, {\rm s}^{2}$  \cite{Mott13,Nish15}.

\revision{The input/output power $P_j$ driving a machine can be subject to strong fluctuations on various time scales induced by renewable resources \cite{Mila13} or consumer behavior \cite{Wood13}. We thus analyze the robustness of the phase-locked state to stochastic fluctuations $\xi_j(t)$, i.e. we set
\be
    P_j(t) = \bar P_j +  \xi_j(t).
\ee}
\revision{The electric power $P_j^{\rm (el)}$ exchanged over a transmission line is determined by the difference of the voltage phase angle of the two terminal nodes. For a common two-pole machine the voltage phase angle equals the mechanical phase angle such that the transmitted real power reads $K_{ij} \sin(\delta_i - \delta_j)$.}
\revision{ The capacity $K_{ij}$ parameter, describing the maximally transmittable power on the transmission line between nodes $i$ and $j$, is determined by the susceptance $B_{ij}$ of the line and the voltage levels $E_i$ and $E_j$ at the two units as $K_{ij}=B_{ij} E_i E_j$.}
\revision{In a complex network of lines and generators the total electric power transmitted from machine $j$ is thus given by $P^{\rm (el)}_j = \sum_i K_{ij} \sin(\delta_i-\delta_j)$. Stable operation of the grid requires all units are in a phase-locked state where $\delta_i-\delta_j$ is fixed. Otherwise, the transmitted electric power $P^{\rm (el)}_j(t)$ would oscillate and average out over time \cite{12powergrid,13powerlong,Mott13,Dorf13,Scha15}. Throughout this Letter we assume a heavily loaded grid, where phase differences are comparably large in the stable phase-locked state. Such a situation is yet uncommon, but will become increasingly likely in the future \cite{Pesc14}. }
\revision{Other scenarios are analyzed in the Supplemental Material \cite{Supplement2017}, including less heavily loaded transmission lines, inverters without inertia $H$ \cite{Simpson-Porco2012} and higher-order power grid models including voltage dynamics \cite{Mach08,Schm16}.}

\begin{figure}[tb]
  \centering
  \includegraphics[width=\columnwidth]{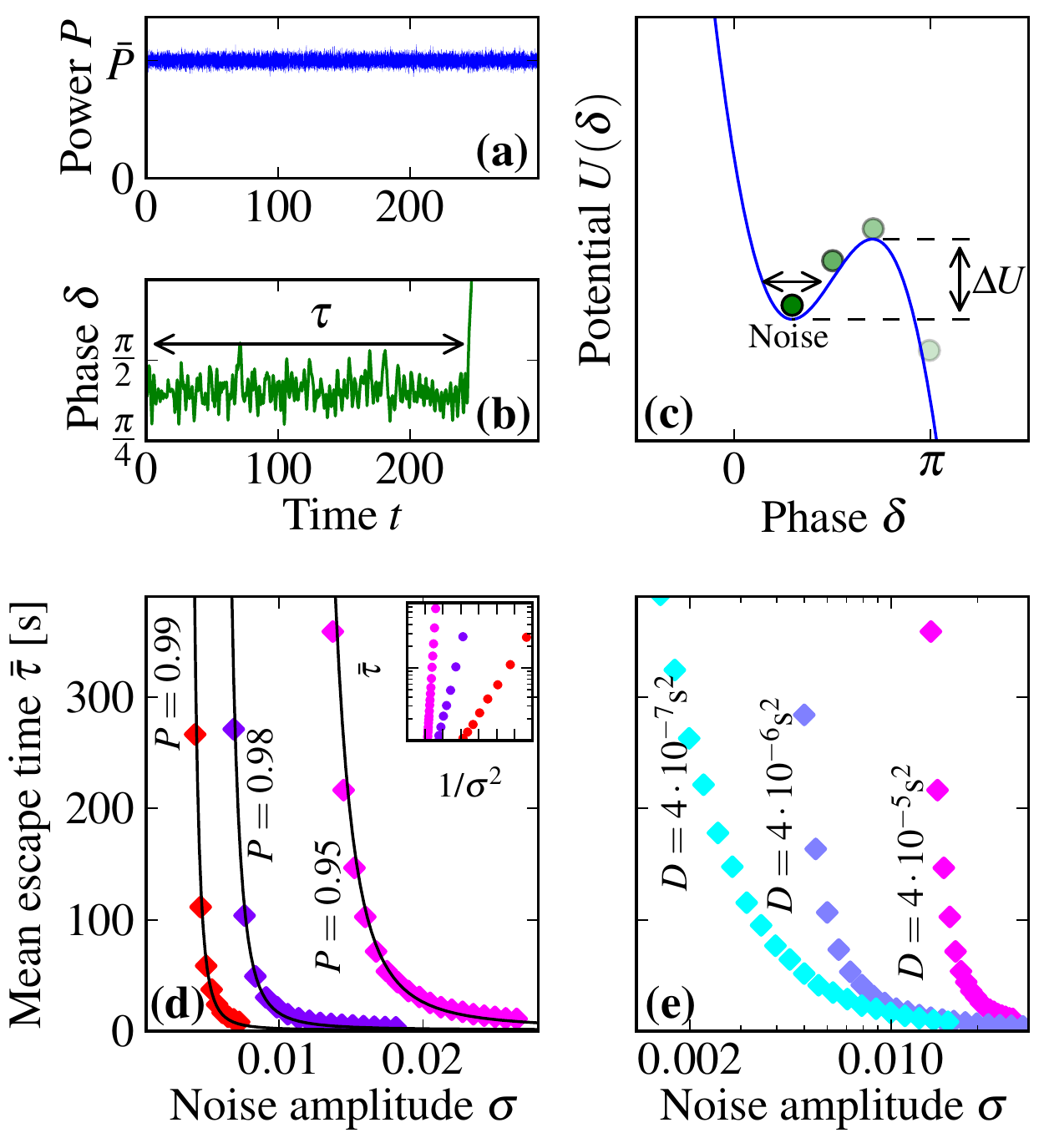}
  \caption{
  \label{fig:kramer}
\revision{Fluctuating power input may desynchronize a synchronous generator. 
(a,b) When the input power $P$ fluctuates,  the generator can lose synchrony to the grid after an escape time $\tau$.   
(c) The generator dynamics corresponds to the motion of a particle in a tilted washboard. The fluctuations can drive the particle over the potential barrier of height $\Delta U$. 
(d) Kramer's escape rate theory predicts the escape process for Gaussian white noise. 
Theoretical prediction, Eq.~(\ref{eq:mean_escape_time}), (black lines) perfectly predicts the mean escape times $\bar \tau$ for intermediate damping ($D=4 \times 10^{-5} \, {\rm s}^{2}$) as checked by direct numerical simulations (symbols). The averaged escape time crucially decreases with increasing load $P$ of the system.
(e) Weaker damping $D$ undermines system robustness, which can become a major problem in future power grids. (Parameters: $H=4 \, \rm s$, $K=1$ and $\bar P  = 0.95$, if not stated otherwise.)}
}
  
\end{figure}

\textit{Robustness of a single generator.--}
\revision{
First, consider a single generator coupled to a large bulk grid. The steady state operation of the generator is described by a stable phase-locked state, a fixed point of the swing equation with a constant phase difference $\delta$ relative to the grid. Fluctuations of the input power $P$ can destabilize the grid as illustrated in Fig. \ref{fig:kramer} (a,b). As soon as the generator leaves the basin of attraction of the stable phase-locked state after some time $\tau$ it rapidly desynchronizes.} \revision{Such a serious contingency requires immediate emergency measures to avoid a global network failure.}

We analyze desynchronization due to white noise by Kramer's escape rate theory \cite{Kramers1940,vKampen,hanggi1990reaction,Gardiner1985} as follows: The equation of motion for the generator is equivalent to a driven dissipative particle moving in a tilted washboard potential \cite{Risken1984},
\revision{i.e.  $\ddot \delta + (\Omega^2 D/H) \dot \delta = (\Omega/2H) \cdot(-\text{d}U/\text{d}\delta + \xi(t))$ with the effective potential \cite{14bifurcation}
\begin{align}
  U(\delta) = -  \bar P \delta - K \cos(\delta).
\end{align}
Thus, to escape the basin of attraction around a local minimum of $U(\delta)$ the particle must overcome a potential barrier of height $\Delta U$ (see Fig. \ref{fig:kramer}(c)), which is determined by the transmitted power $P$ and the capacity $K$. For Gaussian white noise $\xi(t)$ with standard deviation $\sigma$ the mean escape time is given by \cite{vKampen}
\begin{align}
\label{eq:mean_escape_time}
  \bar{\tau} = \frac{2 \pi \lambda}{\sqrt{U''\left(\delta_{\rm min}\right)|U''\left(\delta_{\rm max}\right)|}} \exp{\left( \frac{2 \gamma \, \Delta U}{\sigma^2}\right),}
\end{align}
with effective damping $\gamma = 2 D \Omega$ and $2\lambda= \gamma + \sqrt{\gamma^2 + (8 H/\Omega) |U''\left(\delta_{\rm max}\right)|}$ for intermediate damping and $U''\left(\delta_{\rm min}\right)$ and $U''\left(\delta_{\rm max}\right)$ being the second derivatives of the potential evaluated at the local minimum and the saddle point of the potential $U(\delta)$, respectively \cite{vKampen}. Numerical simulations averaged over 2000 escape processes for each value of $\sigma$ show excellent agreement with this prediction (see Fig. \ref{fig:kramer}(d)).}

\revision{Major concerns about the stability of future power grids arise from the increased transmission needs and lines loads \cite{Pesc14} as well as a possible lack of effective inertia and damping when conventional generators are replaced by renewables \cite{Ulbig2014}. Heavily loaded lines are indeed very susceptible to desynchronization as shown in Fig.~\ref{fig:kramer}(d). Increasing the load $P$ rapidly decreases the escape time $\bar{\tau}$. Similarly, a decrease of the effective damping factor $D$ implies a rapid decrease of $\bar{\tau}$ (Fig.~\ref{fig:kramer}(e) ). In contrast, the inertia $H$ has a minor effect only, as it enters the escape rate (\ref{eq:mean_escape_time}) only through the non-exponential prefactors \cite{Supplement2017}.
}

\textit{Escape dynamics in phase space.--}
The essential factors in Kramer's formula (\ref{eq:mean_escape_time}) are the amplitude of the noise $\sigma$, the effective damping $\gamma$ and the height of the potential barrier $\Delta U$. The theory of random dynamical systems \cite{Gardiner1985} implies that in the limit of weak noise ($\sigma \rightarrow 0$) the system escapes the basin of attraction in the vicinity of a saddle point, where the potential gap to the stable fixed point is smallest. An exemplary escape process in phase space is shown in Fig.~\ref{fig:basin} for the single generator system. Intriguingly, we observe that at any non-zero noise level $\sigma>0$, the trajectory leaves the basin near but not exactly at the saddle point (red dot). \revision{Only in the limit of small perturbations, i.e., $\sigma \rightarrow 0$ does the system leave the fixed point exactly at the saddle.}

\begin{figure}[tb]
  \centering
  \includegraphics[width=\columnwidth]{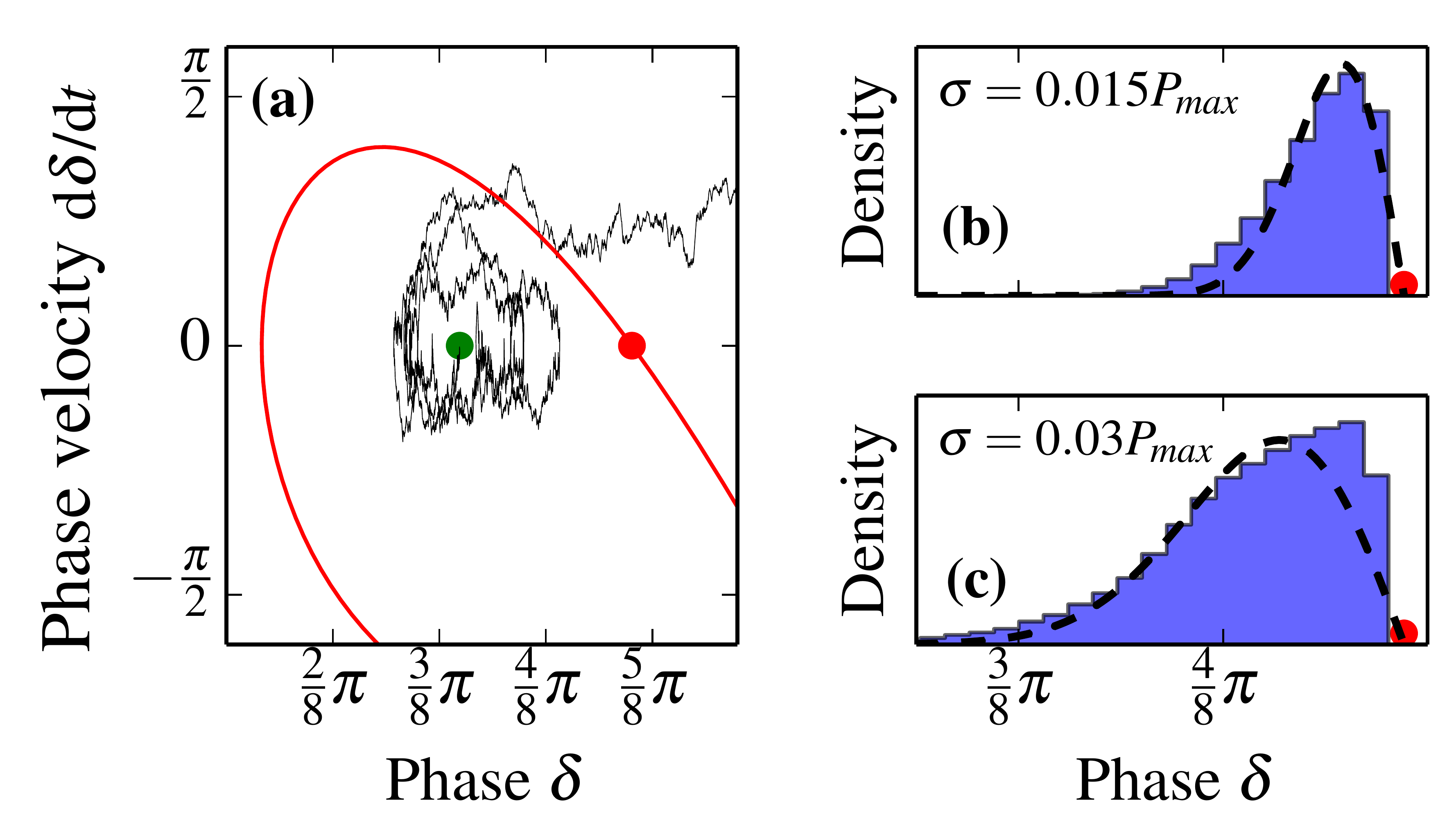}
  \caption{
  \label{fig:basin}
  Where does the system escape?
(a) Exemplary escaping process of the basin of attraction in phase space. The state trajectory (thin black line) crosses the boundary of the basin of attraction (red line) in the vicinity of but not at the saddle point (red disk). The stable fixed point is indicated by a green disk.
(b,c) Probability distribution of exit points, i.e. the crossing points of the trajectory and the basin boundary. Numerical results (histogram)compared to the theoretical prediction, Eq.~(\ref{eq:escape-point}), (dashed black line). With increasing noise the distribution becomes broader, i.e. a crossing at some distance from the saddle point (red disk) becomes more probable. (Parameters: $K=1$ and $\bar P  = 0.95$ as in Fig.~\ref{fig:kramer}.)
}
\end{figure}

The saddle point itself is characterized by a vanishing velocity $d\delta/dt = 0$. Almost all trajectories leave the basin with a non-vanishing velocity $d\delta/dt >0$ ( i.e. `above' the saddle point in the phase space portrait shown in Fig.~\ref{fig:basin}). More precisely, the probability density of the trajectory on the basin boundary in phase space is given by a Weibull function \cite{Schuss} \revision{
\begin{align}
  \label{eq:escape-point}
 p(\delta) = \mathcal{N} \; \delta \exp{\left(-\frac{\lambda^2 \delta^2}{2\sigma^2} - \frac{2\Omega}{H} \cdot \frac{|U''\left(\delta_{\rm max}\right)| \delta^2}{2\sigma^2}\right)} \, ,
\end{align}
}where $\mathcal{N}$ is a normalization constant. This theoretical prediction is equally well confirmed by the numerical simulations as shown in Fig.~\ref{fig:basin} on the right. With increasing noise amplitude the distribution gets broader, i.e. the escape velocity increases.

\begin{figure}[tb]
  \centering
  \includegraphics[width=\columnwidth]{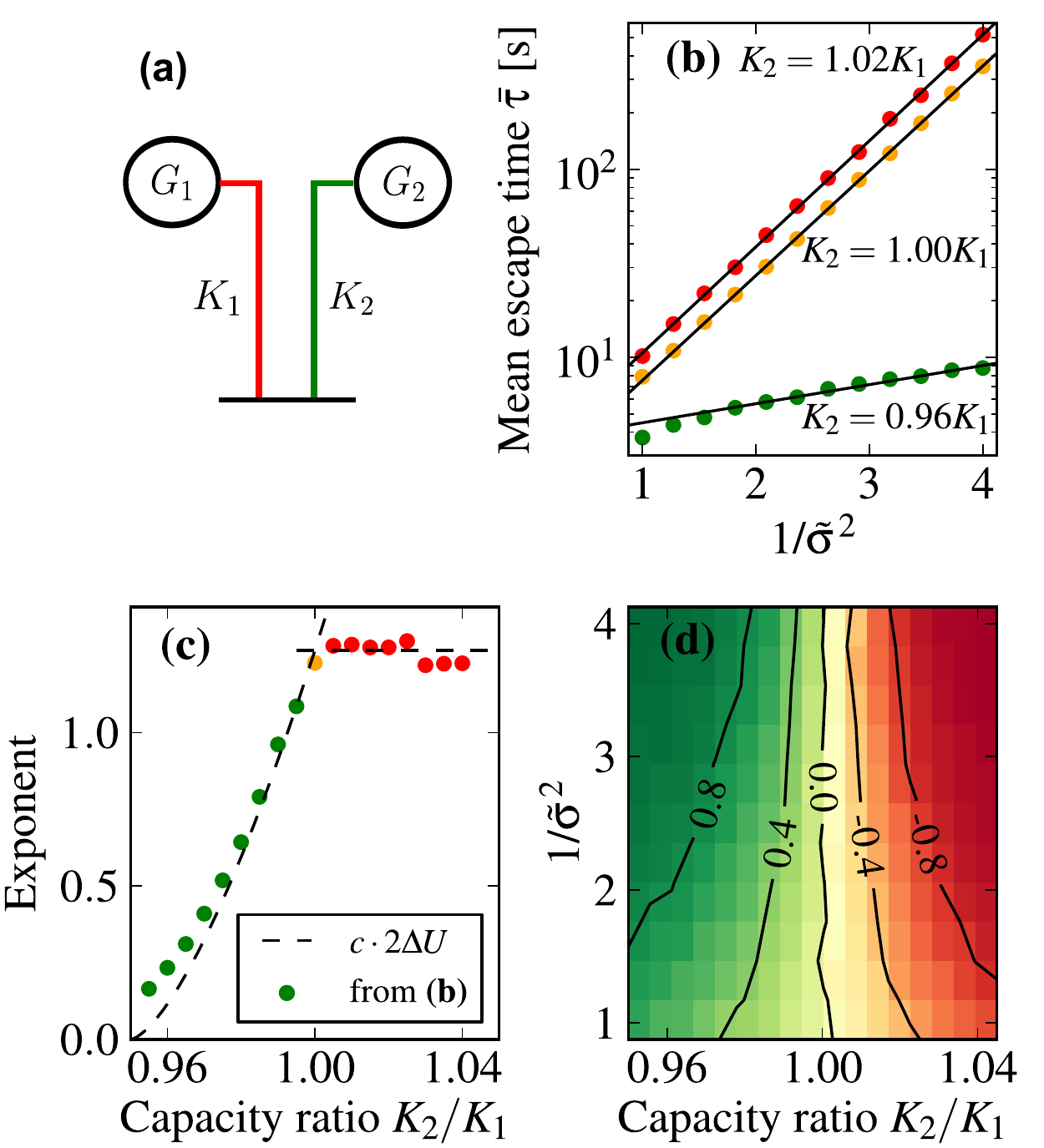}
  \caption{
\label{fig:twogen}  
\revision{The easiest escape route determines the escape time $\bar \tau$. 
(a) Two identical generators are coupled to a third node representing the bulk grid via transmission lines with capacity $K_1=1$ (constant) and $K_2$ (variable), respectively. The transmitted power on both lines is $\bar P_{1,2} = 0.95$. Power fluctuates on all nodes independently.    
(b) The mean escape time $\bar \tau$ as a function of the noise amplitude $\sigma$. Disks represent numerical values, the solid lines are fits to extract the scaling exponent. 
(c) In this scenario, the exponent in Kramer's rate is determined by the lowest barrier $\Delta U$, eq. (\ref{eq:two_dimensional_pot}) of the two-dimensional potential landscape, which is determined by  $\min\{K_1 , K_2\} $, i.e., the \emph{weaker} of the two transmission line capacities. Thus it increases with $K_2$ as long as $K_2 \le K_1$ but depends only on $K_1$ for $K_2 > K_1$. A comparison of numerical results obtained from exponential fits to the data (disks) and the analytical value of the potential barrier $\Delta U$ (with constant $c$) shows very good agreement.
(d) Imbalance of the two escape routes: $p_2-p_1$ with $p_{1,2}$ being the probability that link 1 or 2 is overloaded first, as a function of $K_2/K_1$ and the noise amplitude. For weak noise there is a sharp transition at $K_2/K_1=1$, which smears out for stronger noise. Panels b-d use a re-scaled noise $\tilde{\sigma}=40 \cdot \sigma$.
}}
\end{figure}

\textit{Escape via the weakest link.--}
To maintain a stable operation it is essential to know not only under which conditions, but also how power system operation may become unstable. We first consider a simple system of two identical generators coupled to a bulk power grid with transmission lines of different capacity, both being subject to independent and identically distributed Gaussian white noise (see Fig.~\ref{fig:twogen} (a)). Either of the two generators can become unstable, such that the grid can escape the basin of attraction of the stable phase-locked state via two different routes. The mean escape time is still described by Kramer's rate for intermediate damping, when we take into account that the lower potential barrier along both routes determines the escape (see Fig.~\ref{fig:twogen} (b)). \revision{The two-dimensional potential is then given as \begin{align}
\label{eq:two_dimensional_pot}
  U(\delta_1,\delta_2) = - \bar P_1 \cdot \delta_1 - K_1 \cos(\delta_1) -  \bar P_2 \cdot \delta_2 - K_2 \cos(\delta_2).
\end{align}
}

In the limit of weak noise the escape problem is fully determined by the path with the smallest potential barrier $\Delta U$. To illustrate this we vary the capacity $K_2$ of the transmission line connecting generator $2$ to the bulk grid while the capacity $K_1$ of the other line remains fixed. For $K_2 < K_1$ the robustness is dominated by generator $2$, whose connection is weaker. The exponent in Kramer's formula (\ref{eq:mean_escape_time}) then crucially depends on the value of $K_2$. Indeed, the exponent obtained from the numerical simulations again matches the theoretical predictions well in terms of the potential barrier $\Delta U$ (see Fig.~\ref{fig:twogen} (c)).  If we increase $K_2$ beyond $K_1$, the other transmission line becomes the Achilles' heal of the grid. The potential barrier and hence the exponent in Kramer's formula thus no longer depend on $K_2$. Yet, the non-exponential pre-factor in the formula (\ref{eq:mean_escape_time}) increases by increasing $K_2$ further because the relative transmission line load of the overall system decreases. When the noise becomes stronger, the sharp transition between the two possible escape routes gradually blurs, such that the more strongly connected generator can become unstable too (see Fig.~\ref{fig:twogen} (d)).

\begin{figure}[tb]
  \centering
  \includegraphics[width=\columnwidth]{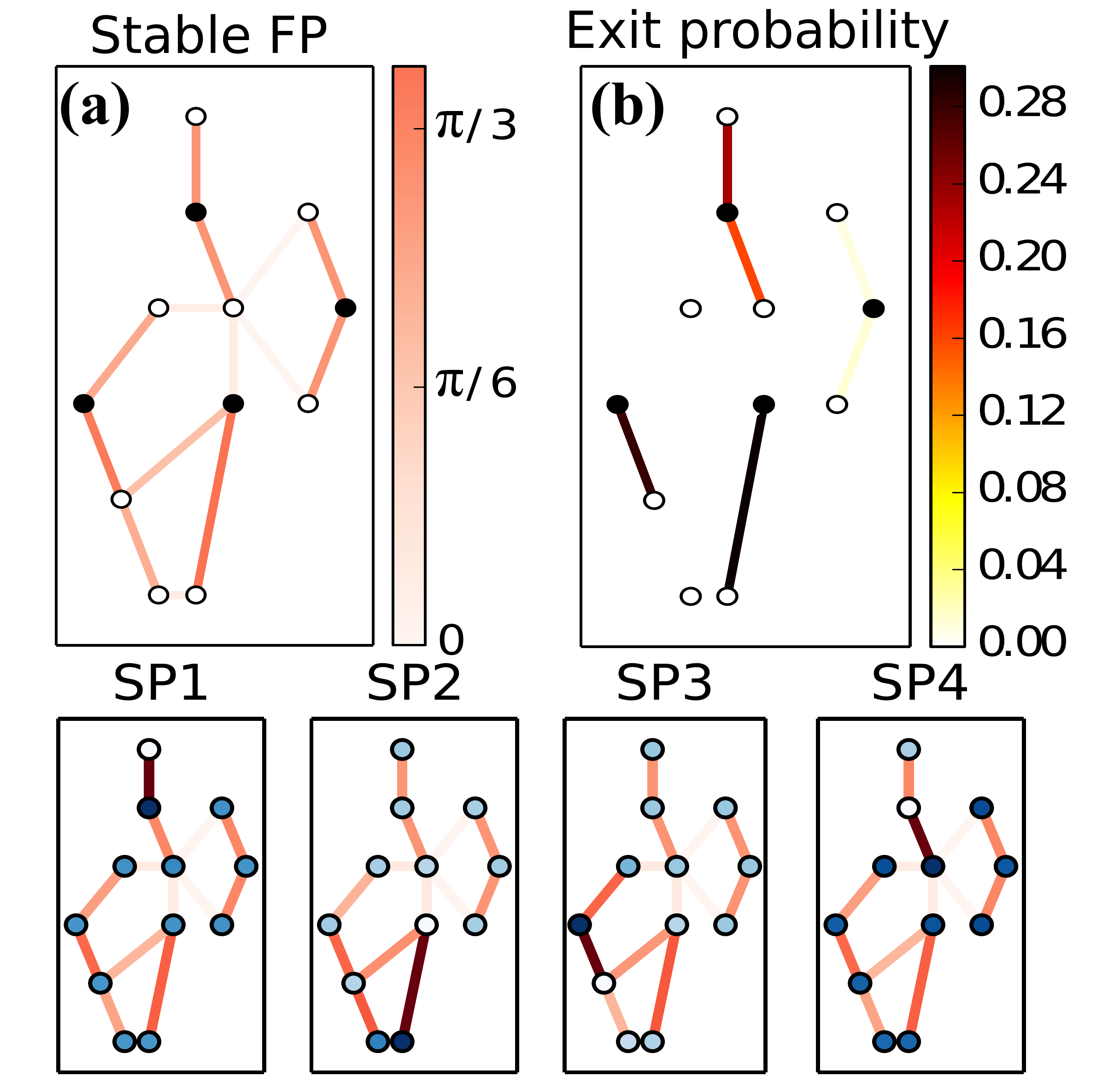}
  \caption{
    \label{fig:saddle}
  Vulnerable links predicted by the topology of saddle points.
  Upper left: The stable phase-locked fixed points in a model power grid with four generators (filled circles) and eight consumers (open circles). Shown is the phase difference $\delta_i - \delta_j$ along the transmission lines.
  Upper right: Probability that a transmission line is overloaded first ($|\delta_i(t) - \delta_j(t)|$ crosses $\pi/2$) when the grid becomes unstable due to a fluctuating power input. Four vulnerable transmission lines are identified. 
  Bottom: The vulnerable transmission lines can be traced back to four different saddle points with comparably low potential barrier. All saddle points have exactly one transmission line (darkest line in each plot) with $|\delta_i - \delta_j| > \pi/2$, corresponding to the vulnerable lines identified in (b). The color scale shows the phase differences as in panel (a).
  The networks consists of four generators ($\bullet$, $P_j =+2P_0$) and eight consumers 
  ($\circ$, $P_j=-P_0$), all lines have capacity $K=24/19 \times P_0$. 
  }
\end{figure}

\textit{Robustness of complex power grids.--}
In power grids with a less simple structure, it is essential to understand how the topology determines the robustness and to identify possible ways of instability. This enables a precise improvement of the grid and the elimination of weak links. Figure \ref{fig:saddle} (a) shows the stable fixed point in a grid with four generator and eight consumer nodes. The consumer dynamics also follows Eq.~(\ref{eqn:swing}), but with $P_j < 0$. A fluctuating input can lead to a loss of synchrony and eventually to a system-wide failure. But where does this instability emerge and which of the transmission lines is most vulnerable?

We simulate the dynamics with all machines subject to independent and identically distributed white noise and record which transmission line becomes overloaded first, i.e. we record for which link $(i,j)$ the phase difference $|\delta_i - \delta_j|$ first crosses $\pi/2$. In this way we identify four transmission lines which are vulnerable. Strikingly, these vulnerable lines are not necessarily the ones which are most heavily loaded in the first place. The loss of synchrony in a complex grid is a collective process, which cannot fully be understood from fundamental properties of single nodes or lines \cite{16critical,Crucitti2004}. 
    
\revision{Instead, Kramer's theory tells that the saddle points of the entire dynamical system are decisive: As above, the grid leaves the basin of attraction of the stable phase-locked state in the vicinity of the saddle points. In a complex network, many saddle points may exist. But for the application of Kramer's theory we only need to consider those saddle points with the lowest potential barrier, as escape through all other saddle points is exponentially suppressed. 
For the system studied here, these saddle points are calculated systematically using a method introduced in \cite{Manik2016}. This method classifies the saddle points by the number of links $(i,j)$ where the phase difference $|\delta_i - \delta_j|$ exceeds $\pi/2$. Typically, the higher this number, the higher is the potential barrier.}

For the sample network depicted in Fig.\ref{fig:saddle} for illustration, this method yields four saddle points with a comparably low potential barrier, all contributing to the escape process (four lower panels). All four saddles have exactly one line where the phase difference $|\delta_i - \delta_j|$ exceeds $\pi/2$. The static analysis thus yields four vulnerable lines which exactly match the lines where overloads have been recorded in the numerical simulations. Even more, Kramer's rate with the respective barrier heights again predicts the exit probabilities (not shown).

\textit{Conclusion. --}
\revision{In this Letter we have analyzed how high-frequency fluctuations impact the dynamical robustness
of electric power grids. Focusing on Gaussian white noise yielded analytical access, thereby providing deeper
insights into the collective dynamics of fluctuation-driven networks. To characterize the robustness of this stochastic system, we derived the scaling of escape times as a function of the grid load, inertia, damping and the noise amplitude. Furthermore, we demonstrated how power networks may escape the regime of stable operation. The grid escapes in the vicinity of (saddle) fixed points with a low potential barrier. Interestingly, these
can typically be assigned to a single overloaded link, thus revealing the weak links of the grid.}

Complementary work on power grid fluctuations \cite{Schm16} addresses the impact of intermittent noise and incorporates features of real wind turbines. Such settings avert the analytic treatment in terms of Kramer's escape
theory. The analytic approach presented in this Letter reveals which factors limit the robustness of power
grid operation to fluctuating inputs. The results may thus not only provide efficient methodology to analyze fluctuation-driven oscillatory systems but may also help planning grid extensions to assure dynamic stability and robustness in future highly renewable power systems.

We gratefully acknowledge support from the Federal Ministry of Education and Research (BMBF grant no.~03SF0472A-E), the Helmholtz Association (via the joint initiative ``Energy System 2050 - A Contribution of the Research Field Energy'' and the grant no.~VH-NG-1025 to D.W.), the G\"ottingen Graduate School for Neurosciences and Molecular Biosciences (DFG Grant GSC 226/2)  to B.S. and the Max Planck Society to M.T.

% --- Literatur -------------------------------------------------------------------

\bibliography{fluct}
%\bibliographystyle{apsrev}

%\newpage
\clearpage
\appendix

%\includepdf[pages=-]{Escape_routes_SI.pdf}
%\includepdf[pages=-,pagecommand={},width=\textwidth]{Escape_routes_SI.pdf}
%\includepdf[pages=-,pagecommand={},width=\textwidth]{Escape_routes_SI.pdf}
%\include{Escape_routes_SI.tex}

%\documentclass[aps,10pt,prl,twocolumn,english,groupedaddress, superscriptaddress]{revtex4-1}
%\usepackage[T1]{fontenc}
%\usepackage[latin9]{inputenc}
%\setcounter{secnumdepth}{3}
%\usepackage{amstext}
%\usepackage{amssymb}
%\usepackage{graphicx}
%\usepackage{hyperref}
%\makeatletter
%\usepackage{babel}
%%%%%%%%%%%%%%%%%%%%%%%%%%%%%% User specified LaTeX commands.
\renewcommand{\thefigure}{S.\arabic{figure}}

\makeatother

\onecolumngrid

%\begin{document}
\begin{center}
\begin{large}
\textbf{Supplemental Material}\\
accompanying the manuscript\\
\textbf{Escape Routes, Weak Links, and Desynchronization in Fluctuation-driven
Networks}
\end{large}
\end{center}

\twocolumngrid

\author{Benjamin Schäfer}

\affiliation{Network Dynamics, Max-Planck-Institute for Dynamics and Self-Organization
(MPI DS), 37077 Göttingen, Germany}

\author{Moritz Matthiae}

\affiliation{Network Dynamics, Max-Planck-Institute for Dynamics and Self-Organization
(MPI DS), 37077 Göttingen, Germany}

\affiliation{Forschungszentrum Jülich, Institute for Energy and Climate Research
- Systems Analysis and Technology Evaluation (IEK-STE), 52428 Jülich,
Germany}

\affiliation{Department of Micro- and Nanotechnology, Technical University of
Denmark, 2800 Kongens Lyngby, Denmark}

\author{Xiaozhu Zhang}

\affiliation{Network Dynamics, Max-Planck-Institute for Dynamics and Self-Organization
(MPI DS), 37077 Göttingen, Germany}

\author{Martin Rohden}

\affiliation{Network Dynamics, Max-Planck-Institute for Dynamics and Self-Organization
(MPI DS), 37077 Göttingen, Germany}

\affiliation{Jacobs University, Department of Physics and Earth Sciences, 28759
Bremen, Germany}

\author{Marc Timme}

\affiliation{Network Dynamics, Max-Planck-Institute for Dynamics and Self-Organization
(MPI DS), 37077 Göttingen, Germany}

\affiliation{Department of Physics, Technical University of Darmstadt, 64289 Darmstadt,
Germany}

\affiliation{Institute for Theoretical Physics, Technical University of Dresden,
01062 Dresden Germany}

\author{Dirk Witthaut }

\affiliation{Forschungszentrum Jülich, Institute for Energy and Climate Research
- Systems Analysis and Technology Evaluation (IEK-STE), 52428 Jülich,
Germany}

\affiliation{Institute for Theoretical Physics, University of Cologne, 50937 Köln,
Germany}

\maketitle

\section{Overview}

Power grids are complex systems and may be described on different
levels of detail \cite{Mach08}. A cornerstone model of power system
dynamics is the swing equation - a second order differential equation
describing the mechanical rotation of synchronous machines. The mechanical
phase angle typically equals the voltage phase angle, such that it
also determines the power flows in the grid. This model is studied
in the main text for heavily loaded lines. In this Supplemental Material
we provide additional arguments why this model is appropriate and
explore different parameter regimes, in particular less heavily loaded
lines and the effects of a decreasing inertia.

Furthermore, we analyze the robustness of different power system models
against noise. Wind turbines and photovoltaics are usually connected
to a grid via power electronic inverters. Power electronic devices
can act as virtual synchronous machines described by the swing equation,
otherwise they should be described as first-order Kuramoto oscillators
\cite{Calabria2015,Simpson-Porco2012,Simpson-Porco2013}. The swing
equation is known to describe the short-term dynamical stability of
a power system. On longer time scales the assumption of a constant
voltage may no longer be satisfied \cite{Auer2016}. In this Supplemental
Material we study grids without inertia in terms of a first-order
model and state the appropriate escape rate formula. In addition,
we investigate the effects of voltage variability numerically in terms
of the third-order model. For simplicity of presentation, we focus
on the scenario of one machine connected to the bulk grid.

\section{Power Grid Models}

First, we define an effective potential $U$ in terms of the voltage
phase $\delta$ as 
\begin{equation}
U\left(\delta\right)=-P\cdot\delta-K\cos\left(\delta\right),\label{eq:potential U}
\end{equation}
with effective power produced $P$ and capacity parameter of a line
$K$, which describes the maximal transmittable power of that line.
In terms of this potential, the swing equation, i.e., the equations
of motion for the voltage phase $\delta$ and its angular velocity
$\omega$ becomes
\begin{eqnarray}
\frac{\text{d}}{\text{d}t}\delta & = & \omega\nonumber \\
\frac{\text{d}}{\text{d}t}\omega & = & -\frac{\Omega^{2}D}{H}\omega+\frac{\Omega}{2H}\left(-\frac{\partial U}{\partial\delta}\left(\delta\right)+\xi\right),\label{eq:swing equation}
\end{eqnarray}
with the reference angular velocity $\Omega=2\pi\cdot50$ Hz, damping
$D$, inertia $H$ and Gaussian white noise $\xi$ with standard deviation
$\sigma$. Parameters used (if not stated otherwise) are $D=4\cdot10^{-5}\text{s}^{2}$,
$H=4\text{s}$ , $P=0.95$, $K=1$. In a fully renewable power grid,
which is dominated by wind and solar power production, there will
be fewer rotating machines than today and the inertia $H$ will be
smaller. In the extreme case all rotating machines are replaced by
inverters feeding wind and solar power into the grid. However, it
has been demonstrated that inverters can be used to act similar to
synchronous machines by providing virtual inertia, see e.g. \cite{DArco2013,schiffer2013synchronization}.
Again, the dynamics of the inverters is preferentially modelled by
using the swing equation (\ref{eq:swing equation}). 

Furthermore, we note that the swing equation (\ref{eq:swing equation})
is also used in coarse grained models \cite{Mach08,Ulbig2014}. In
those models, multiple machines are aggregated into a coherent subgroup.
Each node, representing one sub group, is then described as an oscillatory
machine with effective power $P$, damping $D$ and inertia $H$.
The dynamics is again described by the swing equation.

If we model a system without physical or virtual inertia, each node
is best described as a Kuramoto oscillator \cite{Calabria2015,Simpson-Porco2013,Simpson-Porco2012}
with the equation of motion being\emph{ }
\begin{eqnarray}
\frac{\text{d}}{\text{d}t}\delta & = & \frac{1}{2D\Omega}\left(-\frac{\partial U}{\partial\delta}\left(\delta\right)+\xi\right).
\end{eqnarray}
In the main text we assumed the voltage amplitude to stay constant
even in the scenario of a heavily loaded grid. Here, we consider the
third order model \cite{Schmietendorf2014,Mach08,Auer2016,Ma2016}
which allows the voltage amplitude $E$ to vary over time:
\begin{eqnarray}
\frac{\text{d}}{\text{d}t}\delta & = & \omega\nonumber \\
\frac{\text{d}}{\text{d}t}\omega & = & -\frac{\Omega^{2}D}{H}\omega+\frac{\Omega}{2H}\left(P-EE_{0}B_{0}\sin\left(\delta\right)+\xi\right)\nonumber \\
\frac{\text{d}}{\text{d}t}E & = & \tau_{E}\cdot\left(E_{f}-E+X\left(E_{0}\cos\left(\delta\right)+EB_{11}\right)\right),
\end{eqnarray}
with the voltage time scale $\tau_{E}=2$, bulk voltage $E_{0}=1$,
$E_{f}=1$, $B_{0}=1$, $B_{11}=-\sqrt{1-P^{2}}$ and the voltage
droop $X$. For $X=0$ and $E\left(t=0\right)=1$ the voltage remains
at the fixed point $E^{*}=1$ at all times and reproduces the second
order model while for $X>0$ deviations from the second order model
can be observed. Typical parameter values are taken from \cite{Schmietendorf2014}.

\section{Calculation of $\Delta U$}

In the main text and also in upcoming equations (\ref{eq:2nd order escape time})
and (\ref{eq:1st order escape time}) we use the potential difference
$\Delta U$. It is calculated as follows. First, we determine the
minimum and maximum of the potential $U$ (\ref{eq:potential U})
as 
\begin{eqnarray}
\delta_{\text{min}} & = & \arcsin\left(\frac{P}{K}\right)\\
\delta_{\text{max}} & = & \pi-\arcsin\left(\frac{P}{K}\right)
\end{eqnarray}
respectively. Plugging these into (\ref{eq:potential U}) and calculating
$\Delta U=U\left(\delta_{\text{max}}\right)-U\left(\delta_{\text{min}}\right)$
we obtain 
\begin{equation}
\Delta U=-P\cdot\left(\pi-2\arcsin\left(\frac{P}{K}\right)\right)+2\sqrt{K^{2}-P^{2}}.
\end{equation}

\section{Mean escape time}

We simulate the system consisting of one generator coupled to the
bulk and extract the scaling of the escape time $\bar{\tau}$, depending
on the noise amplitude $\sigma$, the inertia $H$ and the voltage
droop $X$ for the first, second and third order model. For the sake
of consistency and comparability, we define the escape time as that
instance when the system passed $\delta_{\text{crit}}=\pi/2$ and
did not return to the fixed point. Although this does not correspond
exactly to the boundary of the basin of attraction, it enables us
to compare these three different models.

\subsection{2nd Order: Scaling with respect to inertia}

Decreasing inertia $H$ in the swing equation (\ref{eq:swing equation}),
decreases the escape time $\tau$ and thereby the stability of the
system, see fig. \ref{fig:Swing equation effect of inertia}. This
dependency is well-described in Kramer's escape theory
\begin{equation}
\tau_{\text{2nd}}=\frac{2\pi\lambda}{\sqrt{U^{\prime\prime}\left(\delta_{\text{min}}\right)U^{\prime\prime}\left(\delta_{\text{max}}\right)}}\exp\left(\frac{2\gamma\Delta U}{\sigma^{2}}\right)\label{eq:2nd order escape time}
\end{equation}
with 
\begin{equation}
\gamma=2D\Omega
\end{equation}
 and 
\begin{equation}
2\lambda=\gamma+\sqrt{\gamma^{2}+\left(8H/\Omega\right)\left|U^{\prime\prime}\left(\delta_{\text{max}}\right)\right|}
\end{equation}
 as stated in the main manuscript.

\begin{center}
\begin{figure*}
\begin{centering}
\includegraphics[width=2\columnwidth]{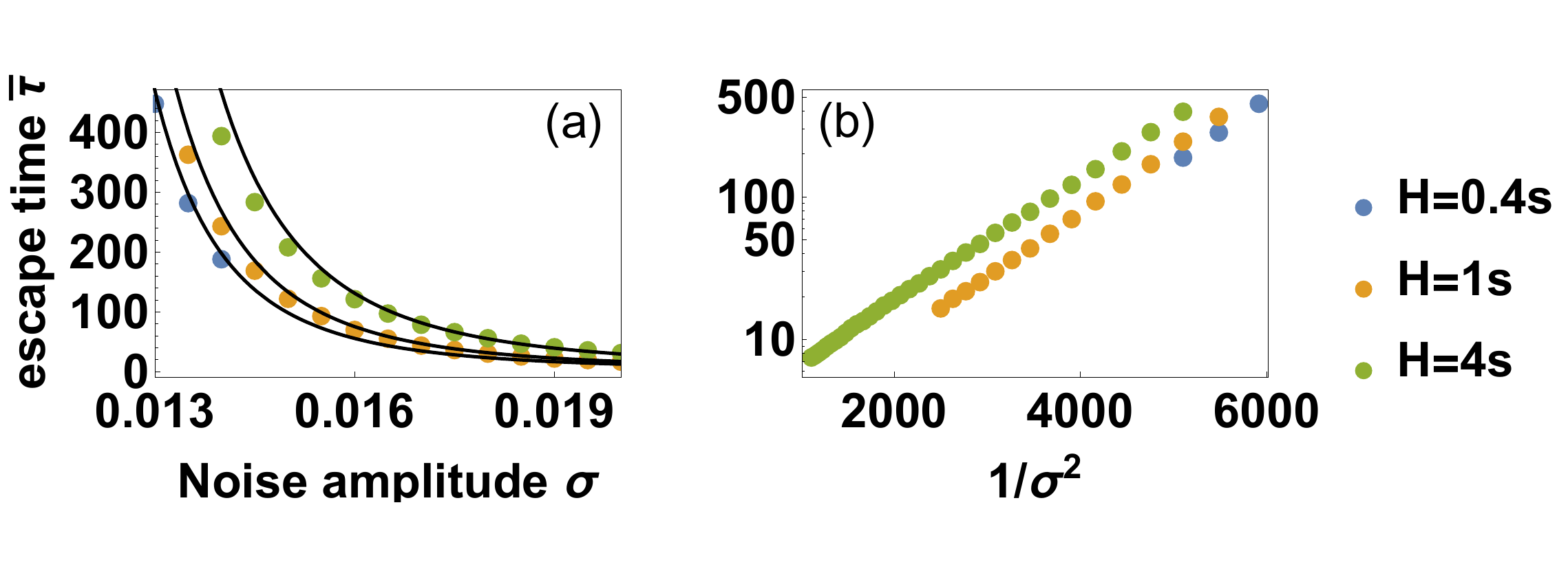}
\par\end{centering}

\caption{The mean escape time $\bar{\tau}$ decreases with decreasing inertia.
We simulated 2000 trials using $D=4\cdot10^{-5}\text{s}^{2}$, $P=0.95$
as parameters for one generator coupled to the bulk grid. Kramer's
formula (black line) shows an excellent agreement with the numerical
results (a). The logarithm of the escape time $\log\left(\bar{\tau}\right)$
is linear in $1/\sigma^{2}$ (b), as predicted by eq. (\ref{eq:2nd order escape time}).
\label{fig:Swing equation effect of inertia}}
\end{figure*}

\par\end{center}

\subsection{2nd Order: Less loaded lines}

\begin{figure*}
\begin{centering}
\includegraphics[width=1.8\columnwidth]{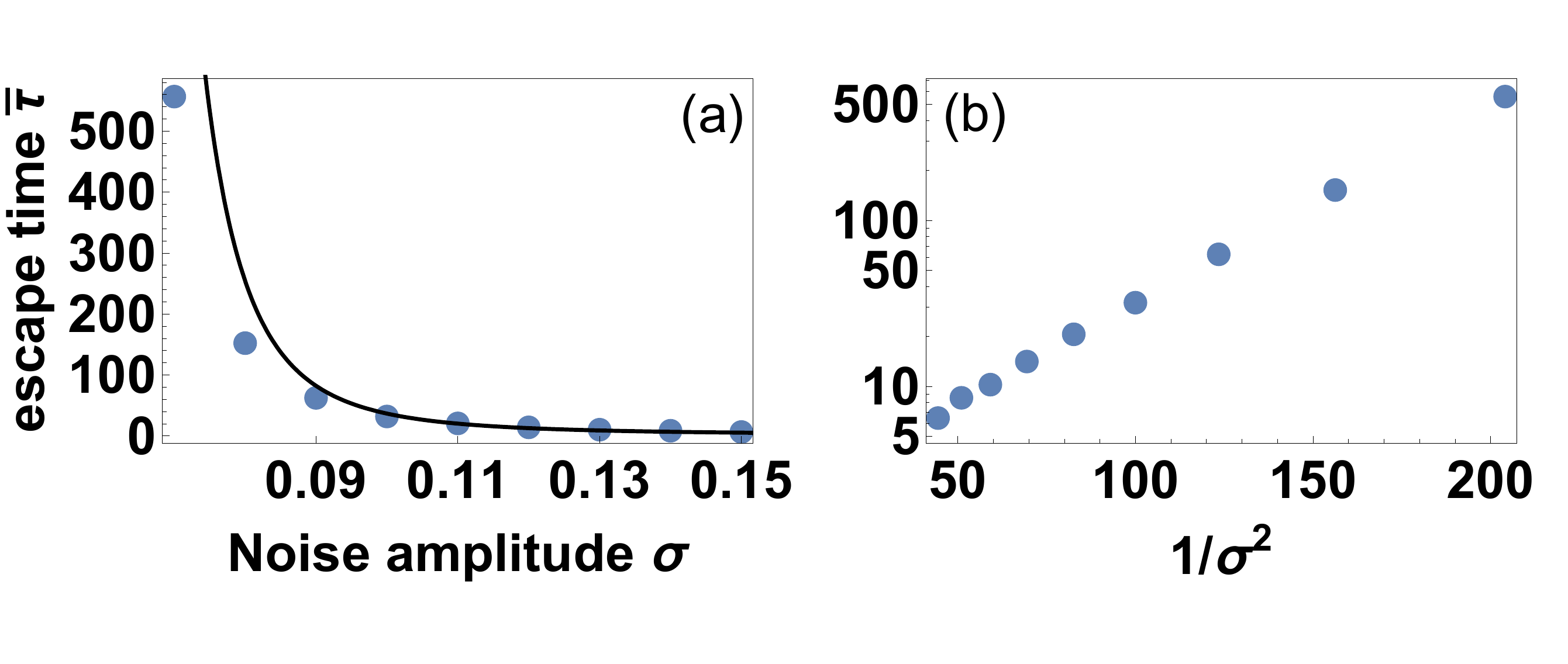}
\par\end{centering}

\caption{Kramer's escape theory also describes less loaded scenarios. We simulated
2000 trials using $H=4\text{s}$, $D=4\cdot10^{-5}\text{s}^{2}$ and
$P=0.5$ as parameters for one generator coupled to the bulk grid.
Kramer's formula (black line) shows an excellent agreement with the
numerical results (a). The logarithm of the escape time $\log\left(\bar{\tau}\right)$
is linear in $1/\sigma^{2}$ (b), as predicted by eq. (\ref{eq:2nd order escape time}).\label{fig:Less loaded scenario}
Note that for similar escape times $\bar{\tau}$ the noise amplitude
has a much larger absolute value compared to more heavily loaded scenarios.}
\end{figure*}

In the main text we only considered highly loaded lines with $P\approx0.95K$,
i.e., the lines were close to maximum load. In fig. \ref{fig:Less loaded scenario}
we display that Kramer's escape theory also holds for less loaded
scenarios. We observe a good agreement of the numerical results and
the analytical prediction by eq. (\ref{eq:2nd order escape time}).
The noise amplitude $\sigma$ needs to be increased significantly
compared to less loaded scenarios, see, e.g. fig. \ref{fig:Swing equation effect of inertia},
to arrive at similar escape times $\bar{\tau}$.

\subsection{First order model}

\begin{figure*}
\begin{centering}
\includegraphics[width=1.8\columnwidth]{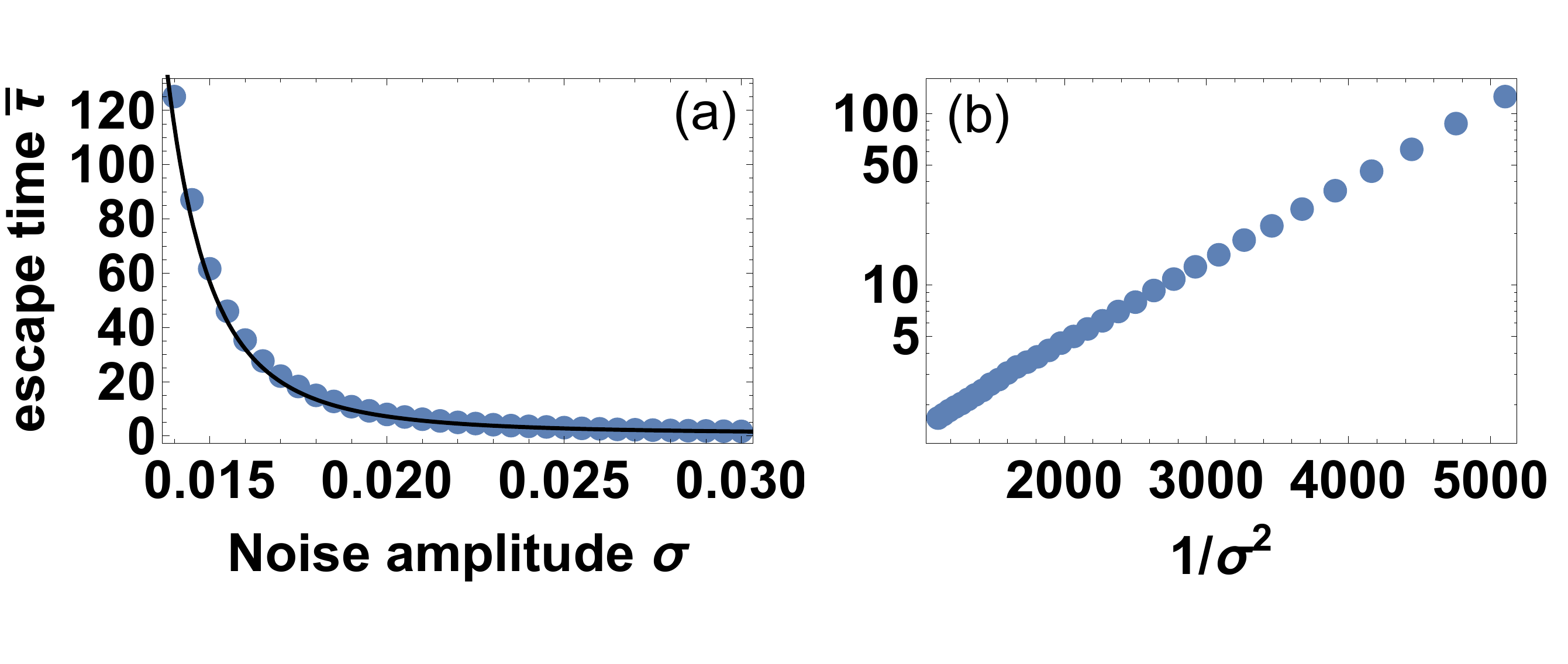}
\par\end{centering}

\caption{The mean escape time $\bar{\tau}$ is well predicted for a first order
model. The black line is Kramer's escape time for the 1st order model,
as given in eq. (\ref{eq:1st order escape time}) and is in excellent
agreement with the numerical data (a). The logarithm of the escape
time $\log\left(\bar{\tau}\right)$ is linear in $1/\sigma^{2}$ (b),
as predicted by eq. (\ref{eq:1st order escape time}). We simulated
2000 trials using $D=4\cdot10^{-5}\text{s}^{2}$ and $P=0.95$ as
parameters for one generator coupled to the bulk grid. \label{fig: First order mean escape plot}}
\end{figure*}

In contrast to the 2nd order model, the 1st order model has only one
globally stable fixed point at $\delta^{*}=\arcsin\left(P\right)$
and we observe transitions from $\delta^{*}$ to $\tilde{\delta^{*}}=n\cdot2\pi+\delta^{*}$
with $n\in\mathbb{Z}$. In a real system, the fast change of the angle
$\delta$ would almost certainly destabilize the system as the power
flow along the line given by $F=K\sin\left(\delta^{*}\right)$ would
change dramatically during the course of this transition and would
most likely violate security regulations \cite{Mach08}. We obtain
Kramer's escape rate for the first order model \cite{vKampen} as
\begin{equation}
\tau_{\text{1st}}=\frac{2\pi}{\sqrt{U^{\prime\prime}\left(\delta_{\text{min}}\right)U^{\prime\prime}\left(\delta_{\text{max}}\right)}}\exp\left(\frac{1}{4D^{2}\Omega^{2}}\cdot\frac{2\Delta U}{\sigma^{2}}\right)\label{eq:1st order escape time}
\end{equation}
and demonstrate perfect agreement of theory and simulations in fig.
\ref{fig: First order mean escape plot}.

\subsection{Third order model}

\begin{figure*}
\begin{centering}
\includegraphics[width=2\columnwidth]{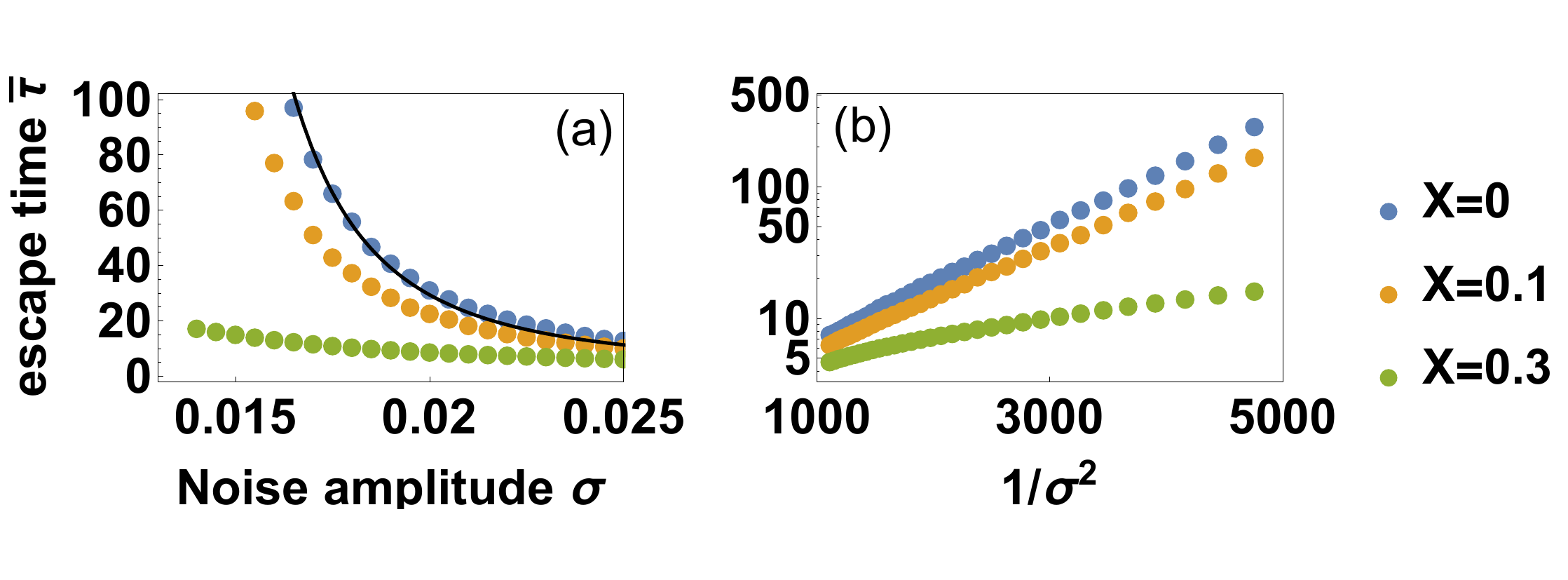}
\par\end{centering}

\caption{The mean escape time $\bar{\tau}$ decreases with increasing voltage
droop $X$. We simulated 2000 trials using $D=4\cdot10^{-5}\text{s}^{2}$,
$P=0.95$, $H=4\,\text{s}$, $\tau_{E}=2$, $E_{0}=1$, $E_{f}=1$
and $B_{11}=-\sqrt{1-P^{2}}$ as parameters for one generator coupled
to the bulk grid. The black line is the prediction based on the 2nd
order (swing equation) model (a) derived in the main text and in eq.
(\ref{eq:2nd order escape time}). \label{fig:Third order escape times}
Note that even with voltage coupling $X>0$, the escape time scales
qualitatively with $\tau\sim\exp\left(c/\sigma^{2}\right)$ , as in
Kramer's theory (b).}
\end{figure*}

The fixed point for the angle $\delta$ in the third order system
is given as $\delta^{*}=\arcsin\left(\frac{P}{E_{0}\cdot E^{*}}\right)$
\cite{Schmietendorf2014}. For sufficiently small values of the voltage
droop $X$, the voltage amplitude $E$ tends to the stable fixed point
$E^{*}=1$. With increasing $X$ this fixed point gets destabilized
and the mean escape time $\bar{\tau}$ of the system decreases. This
effect might change for larger values of $X$. Our theory using Kramer's
escape rate correctly predicts the stability as long as the voltage
changes are sufficiently slow or small, see fig. \ref{fig:Third order escape times}.

%\bibliographystyle{apsrev}
%\bibliography{fluct}

%\end{document}
%

\end{document}